\documentclass[preprintnumbers,prd,twocolumn,showpacs,floatfix,
preprintnumbers,letterpaper,amsmath,amssymb,nofootinbib,superscriptaddress]
{revtex4}
\usepackage{graphicx}
\usepackage{epsfig}
\usepackage{bm}
\usepackage{amsfonts}
\usepackage[latin1]{inputenc}
\usepackage{graphicx}
\usepackage{amssymb}
\usepackage{color}
\usepackage{float}
\usepackage{amsmath}
\usepackage{amsfonts}
\usepackage{dcolumn}
\usepackage{hyperref}
\usepackage{epsfig}
\usepackage{bm}

\newcommand{\Mpl}{M_{\textrm{Pl}}}
\renewcommand{\(}{\left(}
\renewcommand{\)}{\right)}

\newcommand{\nn}{\nonumber}

\begin{document}

\title{Quasi-dilaton non-linear massive gravity: Investigations of background
\\cosmological dynamics}

\author{Radouane Gannouji}
\affiliation{Astrophysics and Cosmology Research Unit, School of
Mathematics,
Statistics and Computer Sciences, Univ. of KwaZulu-Natal, Private Bag
X54001, Durban 4000, South Africa}

 \author{Md. Wali Hossain}
\affiliation{Centre for Theoretical Physics, Jamia Millia Islamia, New
Delhi-110025,
India}

\author{M.~Sami}
\affiliation{Department of Physics, Nagoya University, Nagoya
464-8602, Japan}
\affiliation{Centre for Theoretical Physics,
Jamia Millia Islamia, New Delhi-110025, India}

\author{Emmanuel N. Saridakis}
\affiliation{Physics Division, National Technical University of Athens,
15780 Zografou Campus,  Athens, Greece}
\affiliation{Instituto de F\'{\i}sica, Pontificia Universidad de Cat\'olica
de Valpara\'{\i}so, Casilla 4950, Valpara\'{\i}so, Chile}

\begin{abstract}
We investigate cosmological behavior in the quasi-Dilaton non-linear
massive gravity. We perform a detailed dynamical analysis and
examine the stable late-time solutions relevant to late time
cosmic acceleration. We demonstrate that a dark-energy dominated,
cosmological-constant-like solution is a late attractor of the
dynamics. We also analyze the evolution of the universe at intermediate
times, showing that the observed epoch sequence can be easily
obtained in the model under consideration. Furthermore, we study the
non-singular bounce and the cosmological turnaround which can be
realized for a region of the parameter space.  Last but not least,
we use observational data from Type Ia Supernovae (SNIa), Baryon
Acoustic Oscillations (BAO) and Cosmic Microwave Background (CMB),
in order to constrain the parameters of the theory.
\end{abstract}


\maketitle

\section{Introduction}
It is widely believed that an elementary particle of mass $m$  and
spin $s$ is described by a field which transforms according to a
particular representation of Poincar\'e group. In field theory,
formulated in Minkowski space-time, mass can either be introduced by
hand or via spontaneous symmetry breaking. The spin-2 field
$h_{\mu\nu}$ is supposed to be relevant to gravity as it shares an
important property of universality with Einstein general
relativity(GR) {\it a la Weinberg theorem} and GR can be
thought as an interacting theory of $h_{\mu\nu}$. It is therefore
natural to first formulate the field theory of $h_{\mu\nu}$ in
Minkowski space-time and then extend the concept to non-linear
background.

 The first formulation of a linear massive theory of
gravity was given by Fierz and Pauli \cite{Fierz:1939ix} with a
motivation to write down the consistent relativistic equations for
higher spin fields including spin-2 field. The linear theory of
Fierz-Pauli, however, suffers from van Dam, Veltman, Zakharov (vDVZ)
discontinuity \cite{vanDam:1970vg,Zakharov:1970cc}: in the zero-mass
limit, the theory is at finite difference from GR. The underlying
reason of the (vDVZ) discontinuity is related to the fact the
longitudinal degree of freedom of massive graviton does not decouple
in the said limit; it is rather coupled to the source with the
strength of the universal coupling at par with the massless mode.

It was pointed out by Vainshtein that linear approximation in the
neighborhood of a massive source breaks down below certain distance
dubbed Vainshtein radius and by incorporating non-linearities one
could remove the discontinuity \cite{Vainshtein:1972sx} present in
the linear theory. However, the sixth degree of freedom, suppressed in
Pauli-Fierz theory, which is essentially a ghost known as
Boulware-Deser (BD) ghost \cite{Boulware:1973my}, becomes alive in
the non-linear background.

Efforts were then made to find out the non-linear analog of
Fierz-Pauli mass term requiring the  ghost to be systematically
removed. The authors of Ref.\cite{deRham:2010ik,deRham:2010kj}
discovered a specific nonlinear extension of massive gravity
($dRGT$) that is BD-ghost free in the decoupling limit(see \cite{Hinterbichler:2011tt} for
a review). Subsequently, it was demonstrated \cite{Hassan:2011hr} that the Hamiltonian constraint is
maintained at the non-linear order along with the associated
secondary constraint, which implies the absence of the BD ghost.
Apart from the field theoretic interests of such a construction,
there is a cosmological motivation to study massive gravity. This
formulation provides a new class of (Infra-Red) gravity modification
which can give rise to late-time cosmic acceleration and it is
therefore not surprising that massive gravity has recently generated
enormous interest in cosmology
\cite{Hassan:2011hr,Koyama:2011yg,deRham:2011rn,CuadrosMelgar:2011yw,
Hassan:2011zd,Kluson:2011qe,Gumrukcuoglu:2011ew,
Volkov:2011an,vonStrauss:2011mq,Comelli:2011zm,Hassan:2011ea,
Berezhiani:2011mt,Gumrukcuoglu:2011zh,Khosravi:2011zi,Brihaye:2011aa,
Buchbinder:2012wb,Ahmedov:2012di,Bergshoeff:2012ud,Crisostomi:2012db,
Paulos:2012xe,Hassan:2012qv,Comelli:2012vz,Sbisa:2012zk,Kluson:2012wf,
Tasinato:2012mf,Morand:2012vx,Cardone:2012qq,Baccetti:2012bk,Gratia:2012wt,
Volkov:2012cf,deRham:2012kf,
Berg:2012kn,D'Amico:2012pi,Fasiello:2012rw,
Baccetti:2012ge,Gong:2012yv,Nojiri:2012zu, Deffayet:2012nr,
Chiang:2012vh,Hassan:2012wr,Kuhnel:2012gh,Motohashi:2012jd,Deffayet:2012zc,
Lambiase:2012fv,Gumrukcuoglu:2012wt,Gabadadze:2012tr,Kluson:2012zz,
Tasinato:2012ze,Gong:2012ny,Zhang:2012ap,Park:2012ds,Cai:2012db,
Wyman:2012iw,Burrage:2012ja,Nojiri:2012re,Park:2012cq,Alexandrov:2012yv,
deRham:2012ew,Hinterbichler:2013dv,Langlois:2012hk,Zhang:2013noa,
Haghani:2013eya,DeFelice:2013awa}.

In spite of the  the successes of the framework,  it has challenging
problems to be resolved, namely, the model does not admit spatially
flat FLRW background whereas  isotropic and homogeneous cosmological
solutions ($K=\pm 1$) are perturbatively unstable
\cite{DeFelice:2012mx}. The latter speaks of some inherent
difficulty of $dRGT$. This led the authors of
\cite{D'Amico:2011jj,Gumrukcuoglu:2012aa} to abandon isotropy
whereas in Ref.\cite{Huang:2012pe}, effort was made to address the
problem by making the graviton mass a field dependent quantity through a more
radical approach, namely by extending the theory to a
varying graviton mass driven by a scalar field
\cite{Saridakis:2012jy,Cai:2012ag}. A similar approach was followed
in \cite{D'Amico:2012zv}, where the framework of massive gravity was
extended by introducing a quasi-dilaton field. It was pointed that
the model could rise to a healthy cosmology in the FLRW
background\footnote{ During the finalization of the present work, a
perturbation analysis of the stability of the model was performed in
\cite{DeFelice:2013bxa,D'Amico:2013kya}, where it was shown that the de
Sitter solution of the model may have a ghost instability for short
wavelength modes, shorter than cosmological scales. The calculations were
performed at the linear order of perturbations, or equivalently at second
order in the action. It will be interesting to see whether the ghost exists
in a non-linear approach of the stability analysis.}.

In this work we carry out detailed dynamical investigations of
cosmological behaviors in the aforementioned quasi-Dilaton
non-linear massive gravity. In particular, we perform dynamical
analysis relevant to late time cosmic acceleration, and moreover
we examine the realization of non-singular bouncing solutions. Finally, we
use observational data in order to constrain the model parameters.

The plan of the paper is as follows: In section \ref{model} we
review the quasi-dilaton non-linear massive gravity and its
cosmological implications. In section \ref{Dynamicalbehavior}, we
perform a detailed dynamical analysis and obtain the late-time
stable solutions, while in section \ref{Cosmological evolution} we
investigate the universe evolution  focusing on  bouncing and
turnaround solutions. In section \ref{observations} we use
observations in order to constrain the model parameters and finally
in section \ref{Conclusions}, we summarize the results of our
analysis.

\section{Quasi-Dilaton non-linear massive gravity and cosmology}
\label{model}

In this section we briefly review the quasi-dilaton non-linear
massive gravitational formulation following \cite{D'Amico:2012zv}.
In the first subsection we present the gravitational theory itself,
while in the next subsection we examine its cosmological
implications in presence of matter and radiation.

\subsection{Massive gravity with quasi-dilaton}

Let us consider the following action of non-linear massive gravity
with dilaton
\begin{align}
\label{actiongrav}
\mathcal{S}_{\rm{gr}}=\mathcal{S}_{\rm{EH}}+\mathcal{S}_{\rm{mass}}+
\mathcal{S}_\sigma,
\end{align}
with
\begin{align}
\label{actionEH}
\mathcal{S}_{\rm{EH}}&=\frac{M_{Pl}^2}{2}\int {\rm d}^4 x\sqrt{-g} ~ R
\end{align}
the usual Einstein-Hilbert action,
\begin{align}
\label{actionmass}
\mathcal{S}_{\rm{mass}}&=\frac{m^2 M_{Pl}^2}{8}\int {\rm d}^4 x\sqrt{-g}~
\Bigl[U_2+\alpha_3 U_3+\alpha_4 U_4\Bigr]
\end{align}
the action corresponding to non-linear massive term, along with the
action of a massless dilaton $\sigma$
 \begin{align}
\label{actiondila}
\mathcal{S}_\sigma&=-\frac{\omega}{2}\int{\rm d}^4 x ~(\partial
\sigma)^2.
\end{align}

In action (\ref{actionmass}), $\alpha_3$, $\alpha_4$ are two
arbitrary parameters and $U_i$ are specific polynomials of the
matrix
\begin{equation}
\label{Kdef}
\mathcal{K}^\mu_\nu=\delta^\mu_\nu-e^{\sigma/M_{Pl}}\sqrt{g^{\mu\alpha}
\partial_\alpha\phi^a\partial_\nu\phi^b\eta_{ab}},
\end{equation}
given by
\begin{align}
U_2 &= 4([\mathcal{K}]^2-[\mathcal{K}^2])\\
U_3 &=[\mathcal{K}]^3-3 [\mathcal{K}] [\mathcal{K}^2]+2 [\mathcal{K}^3]\\
U_4 &= [\mathcal{K}]^4-6 [\mathcal{K}]^2 [\mathcal{K}^2]+3
[\mathcal{K}^2]^2+8 [\mathcal{K}] [\mathcal{K}^3]-6 [\mathcal{K}^4].
\end{align}
In  (\ref{Kdef}), $\eta_{ab}$ (Minkowski metric) is a fiducial
metric, and  $\phi^a(x)$ are the St\"{u}ckelberg scalars introduced
to restore general covariance \cite{ArkaniHamed:2002sp}.

\subsection{Cosmology}

In order to apply the above gravitational theory in cosmological
frameworks, we also have to incorporate the matter and radiations
sectors. Thus, the total action we consider is as follows
\begin{align}
\label{action1}
\mathcal{S}=\mathcal{S}_{\rm{EH}}+\mathcal{S}_{\rm{mass}}+
\mathcal{S}_\sigma +\mathcal{S}_m+\mathcal{S}_r,
\end{align}
in which  $\mathcal{S}_m$ and  $\mathcal{S}_r$ denote the
standard matter and radiation actions, corresponding to ideal fluids with
energy densities $\rho_m$, $\rho_r$ and pressures $p_m$, $p_r$ respectively.
In summary, we consider a scenario of usual nonlinear massive
gravity \cite{deRham:2010ik,deRham:2010kj} coupled with a dilaton field
$\sigma$, where the coupling is introduced through (\ref{Kdef})
\cite{D'Amico:2012zv}.

Next, we consider a flat  Friedmann-Lema\^itre-Robertson-Walker (FLRW) metric
of
the form
\begin{align}
{\rm d}s^2=g_{\mu\nu}d x^\mu d x^ \nu=-N(t)^2dt^2+a^2(t)\delta_{ij} dx^i
dx^j,
\end{align}
while for the  St\"{u}ckelberg scalars we consider the ansatz
\begin{align}
\phi^0=f(t), ~~~ \phi^i=x^i.
\end{align}
In this case the actions (\ref{actionEH})-(\ref{actiondila}) become:
\begin{align}
\mathcal{S}_{\rm{EH}}&=-3M_{Pl}^2\int {\rm d}t \Bigl[\frac{a\dot
a^2}{N}\Bigr]\\
\label{gravitonpart}
\mathcal{S}_{\rm{mass}}&=3m^2 M_{Pl}^2\int {\rm d}t a^3\Bigl[N G_1(\xi)-\dot
f a G_2(\xi)\Bigr]\\
\mathcal{S}_\sigma&=\frac{\omega}{2}\int{\rm d}t~ a^3 \Bigl[ \frac{\dot
\sigma^2}{N}\Bigr],
\label{actiondilaton}
\end{align}
where we have defined
\begin{align}
\label{G1def}
G_1(\xi) &= (1-\xi)\Bigl[2-\xi+\frac{\alpha_3}{4} (1-\xi)(4-\xi)+\alpha_4
(1-\xi)^2\Bigr]\\
\label{G2def}
G_2(\xi) &= \xi (1-\xi)\Bigl[1+\frac{3}{4}\alpha_3 (1-\xi)+\alpha_4
(1-\xi)^2\Bigr]
\end{align}
being functions of the introduced dilaton-variable
\begin{equation}
\xi=\frac{e^{\sigma/M_{Pl}}}{a}.
\label{eq:xi}
\end{equation}

Variation with respect to $f$  gives the constraint equation (as an
essential feature of any model of non-linear massive gravity)
\begin{align}
\label{eq:Constraint}
G_2(\xi)=\frac{C}{a^4},
\end{align}
where $C$ is a constant of integration.

Variation with respect to the lapse function $N$ and setting $N=1$
at the end, leads to the first Friedmann equation
\begin{align}
\label{eq:Fried}
3M_{Pl}^2H^2=\frac{\rho_m+\rho_r-3 m^2 M_{Pl}^2G_1}{1- \frac{\omega}{6}
\Bigl(1-4\frac{G_2}{\xi G_2'}\Bigr)^2},
\end{align}
where primes in $G$'s denote derivatives with respect to their argument
$\xi$. Similarly, variation with respect to $\sigma$ provides the dilaton
evolution equation
 \begin{equation}
 \frac{\omega}{a^3}\frac{d}{Ndt}(a^3\dot{\sigma})+3\Mpl
m^2\left[F_1(\xi)+a\dot{f}F_2(\xi)\right]=0,
\end{equation}
where
\begin{eqnarray}
 F_1(\xi) &&
=3(1+\frac{3}{4}\alpha_3+\alpha_4)\xi-2(1+\frac{3}{2}
\alpha_3+3\alpha_4)\xi^2 \nn \\&&
 +\frac{3}{4}(\alpha_3+4\alpha_4)\xi^3 \\
 F_2(\xi) && =
(1+\frac{3}{4}\alpha_3+\alpha_4)\xi-2(1+\frac{3}{2}\alpha_3+3\alpha_4)\xi^2
\nn \\&&
 +\frac{9}{4}(\alpha_3+4\alpha_4)\xi^3-4\alpha_4\xi^4.
 \end{eqnarray}

Finally, the evolution equations close by considering the continuity
equations for matter and radiation, namely $\dot{\rho}_m
+3H(\rho_m+p_m)=0$ and $\dot{\rho}_r +3H(\rho_r+p_r)=0$
respectively. 
 
Before proceeding to the detailed analysis of the model, let us make a few
comments. Firstly, as it was first noticed for a similar model in
\cite{Hinterbichler:2013dv}, the Friedmann equation (\ref{eq:Fried}) can
exhibit singularities when the denominator goes to zero. However, as we will
show in the next section, these singularities separate the phase space in 
disconnected branches, that is the universe cannot evolve from one
to the other.

As a second remark we mention that in the case where $C\neq 0$, in the
constraint equation (\ref{eq:Constraint}) it appears a function of
$1/a^4$ which plays, in the early Universe, a role similar to ``Dark
Matter'' or ``Dark
Radiation''. In particular, when $\alpha_4 \neq 0$, in
the early Universe ($a \rightarrow 0$) relation (\ref{eq:xi}) gives
$\xi \rightarrow \infty$. This implies $G_2\simeq -\alpha_4 \xi^4=C/a^4$
leading the Friedmann equation  (\ref{eq:Fried}) to be
\begin{align}
3M_{Pl}^2 H^2 \simeq  \rho_m+\rho_r+3 m^2 M_{Pl}^2
\left(\alpha_4+\frac{\alpha_3}{4}\right)\Bigl(-\frac{C}{\alpha_4}\Bigr)^{3/4}
a^ { -3 },
\end{align}
that is an effective dark-matter term appears.
Similarly, when $\alpha_4=0$
we acquire
\begin{align}
3\Bigl(1-\frac{\omega}{54}\Bigr)M_{Pl}^2 H^2\simeq \rho_m+\rho_r+m^2
M_{Pl}^2C a^{-4},
\end{align}
that is we get an effective dark radiation term.

Finally, we mention that in the particular case $C=0$, 
(\ref{eq:Constraint}) leads to a constant $\xi$, and the Friedmann
equation  (\ref{eq:Fried}) becomes
\begin{align}
3M_{Pl}^2 H^2 = \rho_m+\rho_r-\alpha_\Lambda +\beta H^2\,,
\end{align}
where $\alpha_\Lambda, \beta$ are constants. It is interesting to notice that
this relation appears as a particular case of
Holographic models where the IR cut-off is fixed at the Hubble scale. In fact
in the aforementioned  theories the number of degrees of freedom in a finite
volume should be finite and related to the area of its boundary. This gives
an upper bound on the entropy contained in the visible Universe. Following an
idea suggested by \cite{Cohen:1998zx,Horava:2000tb} where a long distance
cut-off is related to a short distance cut-off, in \cite{Hsu:2004ri} it was
suggested
to take the largest distance as the Hubble scale or the event horizon. Later
these models were dubbed holographic dark energy \cite{Li:2004rb}.

All these particular cases reveal the richness and the capabilities of the
scenario.  In what follows we shall cast the evolution equations
in the autonomous form and investigate their implications for late
time cosmology.

\section{Dynamical behavior}
   \label{Dynamicalbehavior}

In this section we perform a dynamical analysis of the scenario at
hand, which allows us to bypass the complexities of the equations
and obtain information for the late-time, asymptotic, cosmological
behavior \cite{Copeland:1997et,Ferreira:1997au,Chen:2008ft}. We
mention here that in \cite{D'Amico:2012zv} the authors investigated
the cosmological implications in a specific setting allowing them to
obtain simple analytical solutions. However, in this work we are
interested in the full dynamical investigations of evolution
equations.

\subsection{Phase-space analysis}

In order to perform a phase-space analysis, we first transform the
involved cosmological equations into their autonomous form
introducing  suitable auxiliary variables. We extract the critical
points of the dynamical system under consideration,  perturb the
system around them and  analyze their  stability by examining the
eigenvalues of the corresponding perturbation matrix.

Let us first introduce the density parameters
\begin{align}
\Omega_m &=\frac{\rho_m}{3 M_{Pl}^2 H^2}\\
\Omega_r &=\frac{\rho_r}{3 M_{Pl}^2 H^2}\\
\label{OmegaLambda}
\Omega_{\Lambda} &=-m^2 \frac{G_1}{H^2}\\
\Omega_\sigma &=\frac{\omega}{6} \Bigl(1-4 \frac{G_2}{\xi G_2'}\Bigr)^2,
\end{align}
with which we can re-write the Friedmann equation (\ref{eq:Fried}) as
 \begin{align}
\label{Fr1b}
\Omega_m+\Omega_r+ \Omega_\Lambda+\Omega_\sigma=1.
\end{align}
Thus, we can now transform the above cosmological system into its
autonomous form, using only the dimensionless variables $\Omega_r$,
$\Omega_\Lambda$ and $\xi$, while (\ref{Fr1b}) will be used to eliminate
$\Omega_m$. Doing so we obtain
\begin{align}
\label{eqr}
\frac{{\rm d}\Omega_r}{{\rm d}\ln a} &= -2\Omega_r \Bigl(2+\frac{\dot
H}{H^2}\Bigr)\\
\label{eqLambda}
\frac{{\rm d}\Omega_\Lambda}{{\rm d}\ln a} &=-2\Omega_\Lambda
\Big(2\frac{G_2 G_1'}{G_1 G_2'}+\frac{\dot H}{H^2}\Bigr)\\
\label{eqxi}
\frac{{\rm d}\xi}{{\rm d}\ln a} &=-4\frac{G_2}{G_2'},
\end{align}
where the combination $\frac{\dot
H}{H^2}$ can be acquired differentiating the first Friedmann equation
(\ref{eq:Fried}) as
 \begin{align}
\frac{\dot
H}{H^2}=\frac{-9\Omega_m-12\Omega_r-12\frac{G_2}{G_2'}\Bigl[\frac{G_1'}{G_1
}\Omega_\Lambda+\frac{\omega}{6}\frac{\rm d}{{\rm d}\xi}(1-4\frac{G_2}{\xi
G_2'})^2\Bigr]}{6-\omega \Bigl[1-4\frac{G_2}{\xi G_2'}\Bigr]^2}.
\label{HdotH2}
\end{align}

A comment about the role of dilaton in the scenario under
consideration is in order. The dilaton role is twofold, first, it
has an effect through its impact on the graviton mass, since in the
above equations $m$ appears multiplied by a function of $\xi$, that
is of the exponential of the dilaton field. This is what we call
$\Omega_\Lambda$. Secondly, the dilaton has an effect through its
kinetic term in (\ref{actiondilaton}), which is of course
proportional to the parameter $\omega$. This is what we call
$\Omega_\sigma$. Therefore, strictly speaking, in the present
scenario the effective dark energy sector includes both
$\Omega_\Lambda$ and  $\Omega_\sigma$, since these two contributions
cannot be distinguished observationally (we write them separately
only for clarity). Thus:
 \begin{align}
\label{OmDE}
\Omega_{DE}\equiv \Omega_\Lambda+\Omega_\sigma=1- \Omega_m-\Omega_r.
\end{align}

Finally, note that in terms of the auxiliary variables $\Omega_r$,
$\Omega_\Lambda$ and $\xi$, we can express the physically
interesting observables  such as the total (effective)
equation-of-state parameter $w_{eff}$, the deceleration parameter
$q$ and the dark energy equation-of-state parameter $w_{DE}$
respectively as
\begin{equation}
w_{eff}=-1-\frac{2}{3}\frac{\dot
H}{H^2}
\label{Weff}
\end{equation}
\begin{align}
\label{qcase1}
q=-1-\frac{\dot
H}{H^2}=\frac{1}{2}+\frac{3}{2}w_{eff}.
\end{align}
\begin{equation}
w_{DE}=\frac{w_{eff}-w_m\Omega_m-w_r\Omega_r}{\Omega_\Lambda+
\Omega_\sigma},
\label{WDE}
\end{equation}
where the combination $\dot
H/H^2$ is given by (\ref{HdotH2}). Lastly, in the following for simplicity
we focus on the usual cases of standard dust matter $w_m=0$ and standard
radiation
$w_r=1/3$.

The critical points of the above autonomous system are extracted setting
the left hand sides of equations (\ref{eqr})-(\ref{eqxi}) to zero. In
particular, equation (\ref{eqxi}) implies that at the critical points
$G_2(\xi)=0$, which according
to (\ref{G2def}) gives $\xi=0,\xi=1$ and
\begin{align}
\label{xipm}
\xi_\pm &=1+\frac{3\alpha_3\pm \sqrt{9\alpha_3^2-64\alpha_4}}{8\alpha_4}.
\end{align}
In order for the critical points to be physical they have to possess
$0\leq\Omega_r\leq1$, $0\leq\Omega_\Lambda+\Omega_\sigma\leq1$, $0\leq\xi$
and of course $\xi\in {\cal{R}}$.
Now, in order for
$\xi_\pm\in {\cal{R}}$ we require $9\alpha_3^2-64\alpha_4\geq0$ while
$0\leq\xi$ leads to the necessary conditions for $\alpha$'s using
(\ref{xipm}).

\begin{table*}[ht]
\begin{center}
\begin{tabular}{|c|c|c|c|c|c|c|c|c|c|c|c|}
\hline Cr.P.&
$\Omega_r$ & $\Omega_\Lambda$ & $\xi$ & $\Omega_m$ &  $\Omega_\sigma$ &$q$&
$w_{\rm eff}$ & $w_{DE}$ &Existence conditions& Stability & Eigenvalues\\
\hline
A&0 & 0  & 0  & $1-3\omega/2$ & $3\omega/2$ &$1/2$   &  0 & 0
&$0\leq\omega\leq2/3$& Saddle point & -4, 3, -1 \\
\hline
B&0 & $1-3\omega/2$  & 0 & 0 & $3\omega/2$ &$-1$   &  -1 & -1
&$0\leq\omega\leq2/3$&  Attractor  & -4, -4, -3 \\
 \hline
C&$1-3\omega/2$ & 0  &0  & 0 & $3\omega/2$ &1& $1/3$ & $1/3$
&$0\leq\omega\leq2/3$& Saddle point & -4, 4, 1 \\
 \hline
D&0 & 0  &1  & $1-\omega/6$ & $\omega/6$ &$1/2$   &  0 & 0
&$0\leq\omega\leq6$ & Attractor &  -4, -1, -1\\
 \hline
$E_\pm$&0 & 0  & $\xi_\pm$ & $1-\omega/6$ & $\omega/6$ &$1/2$& 0 & 0 &
$0\leq\omega\leq6$,  $0\leq\xi_\pm$, $\xi_\pm\in {\cal{R}}$ & Saddle point
& -4,-1, 3\\
 \hline
$F_\pm$&$1-\omega/6$ & 0  & $\xi_\pm$ & 0 & $\omega/6$ &1& $1/3$ & 1/3
&$0\leq\omega\leq6$,  $0\leq\xi_\pm$, $\xi_\pm\in {\cal{R}}$& Saddle point
& -4,4,1\\
\hline
$G_\pm$&0 & $1-\omega/6$  & $\xi_\pm$ & 0 & $\omega/6$  &$-1$&-1 & -1
&$0\leq\omega\leq6$,  $0\leq\xi_\pm$, $\xi_\pm\in {\cal{R}}$& Attractor  &
-4,-4,-3 \\
 \hline
$I$&$\Omega_r$ & $1-\omega/6-\Omega_r$  & $1$ & 0& $\omega/6$  &$1$&$1/3$
&  $1/3$
&$0\leq\omega\leq6$&  Saddle  & -4,1,0
 \\
 \hline
\end{tabular}
\end{center}
\caption[crit]{\label{stability1}  The real and physically meaningful
critical points of the autonomous system (\ref{eqr})-(\ref{eqxi}), their
existence conditions, the eigenvalues of the perturbation matrix, and the
deduced stability conditions. We also present the corresponding
values of the various density parameters, and the values of the
observables: deceleration parameter $q$, total equation-of-state
parameter $w_{\rm eff}$ and dark energy equation-of-state
parameter $w_{DE}$. }
\end{table*}



The real and physically meaningful critical points
($\Omega_r,\Omega_\Lambda,\xi$) of the autonomous system
(\ref{eqr})-(\ref{eqxi}) are presented in Table \ref{stability1}, along
with their existence conditions. We mention here that there is an
additional existence condition, namely the obtained $H^2$ from
(\ref{eq:Fried}) must be always positive, a condition which constrains the
allowed $\alpha_3$ and $\alpha_4$ values. We will come back to this later
on.

For each critical point we
calculate the $3\times3$ perturbation matrix of the corresponding
linearized perturbation equations, and  examining the sign of the real part
of its eigenvalues we determine the type and stability of this point
(positive real parts correspond to unstable point, negative real parts
correspond to stable point, and real parts with different signs correspond
to saddle point). The eigenvalues and the stability results are summarized
in Table \ref{stability1}.

Finally, for completeness, in the same Table we
present the corresponding values of the rest density parameters  $\Omega_m$
and $\Omega_\sigma$, as well as the calculated  values for the observables
such as the deceleration parameter $q$, the total equation-of-state
parameter $w_{\rm eff}$, and the dark energy equation-of-state
parameter $w_{DE}$. These results
are in agreement with the general analysis of nonlinear massive gravity of
\cite{Leon:2013qh}.

Before proceeding to the discussion of the physical implications of these
results, we make a remark concerning the structure of the phase space. As we
mentioned in the Introduction, the Friedmann equation (\ref{eq:Fried})
exhibits singularities when the denominator goes to zero. In particular,
in the scenario at hand we obtain various singularities localized
at the roots of $1- \frac{\omega}{6}\Bigl(1-4\frac{G_2}{\xi G_2'}\Bigr)^2$,
which, for every
parameter $\omega$, is a function of the rescaled field-variable $\xi$ only.
Since $\omega$ must be relatively small in order to
satisfy observations (this will be confirmed later on), assuming $\omega \ll
1$ the zeros of the previous function nearly coincide with
the roots of $G_2'(\xi)$. But as we mentioned above, the critical points
correspond to $G_2(\xi)=0$. Thus, using the Rolle's theorem, we conclude that
each
cosmological late-time solution ($G_2(\xi)=0$) and the corresponding region,
is disconnected from the others by a
singularity ($G_2'=0$). In other words, during its evolution, the universe
cannot transit from one of these regions to another.

\subsection{Physical implications}

In the previous subsection we performed a complete phase-space analysis of
the scenario at hand, we extracted the late-time stable solutions
(attractors) and we calculated the corresponding observables. Here we
discuss their cosmological implications.

As we observe from Table \ref{stability1}, there exist four stable
critical points, that can attract the universe at late times, namely $B$,
$D$ and the two points $G_\pm$. Points $B$ and $G_\pm$ correspond to a
dark-energy dominated, accelerating universe, where dark energy behaves
like
cosmological constant (de Sitter solutions). These features make these
points very good candidates for the description of late-time universe. On
the other hand, point $D$ corresponds to a non-accelerating universe, with
dark energy and matter density parameters of the same order (that is it
can alleviate the coincidence problem), but the dark energy has a stiff
equation of state. Therefore, this point is disfavored by observations.

Apart from the above stable late-time solutions, the scenario of
quasi-Dilaton non-linear massive gravity possesses the non-accelerating,
stiff dark-energy points $A$,$C$,$E_\pm$,$F_\pm$, and the critical curve $I$.
These points are saddle ($I$ is non-hyperbolic curve with saddle behavior)
and thus they cannot be the late time states of the universe, and moreover
their cosmological features are not favored by observations.

Let us make a comment here on the appearance of more than one stable
critical points, which for some parameter choices can exist
simultaneously. Although the bi-stability and multi-stability is usual in
complicated dynamical systems \cite{PoincareProj,Leon2011}, with each
critical point attracting orbits initially starting inside its basin of
attraction, in the present scenario the case is simpler, since these
points usually lie in the  disconnected regions we
mentioned in the previous subsection,  separated
by the $\xi$ values that make the denominator of (\ref{eq:Fried}) zero,
and thus to Big-Rip-type divergences
\cite{Copeland:2006wr,Bamba:2008ut,Capozziello:2009hc,Saridakis:2009jq}.
Finally, note that for $\omega>6$ there are not stable late-time
solutions, that is the corresponding cosmology is not interesting.

\begin{figure}[ht]
\begin{center}
\mbox{\epsfig{figure=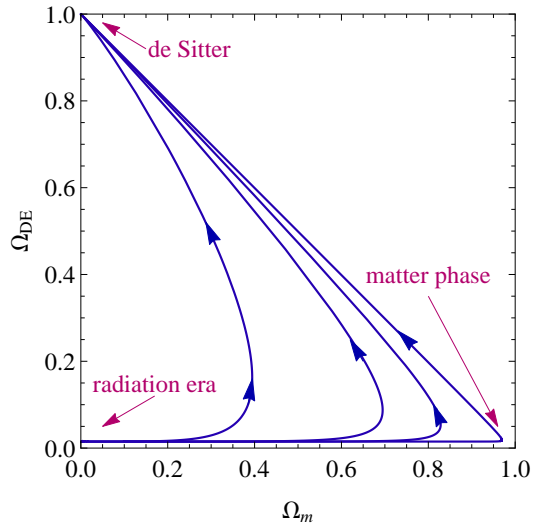,width=7cm,angle=0}}
\caption{{\it
Trajectories in the $\Omega_{DE}$-$\Omega_m$ plane of the
cosmological scenario of quasi-Dilaton non-linear massive gravity, for 
the parameter choice $\alpha_3=1$,$\alpha_4=1$ and $\omega=0.01$. In this
specific example the universe at late times is led to the de Sitter
solution.}} \label{Fig1}
\end{center}
\end{figure}
Finally, as we previously mentioned, the $\alpha_3$-$\alpha_4$ parameter
subspace is constrained by the condition the obtained $H^2$ from
(\ref{eq:Fried}) to be always positive. The corresponding constraints are
complicated and we do not present them here, however they do get a very
simple form when the universe evolves towards its interesting stable
late-time states $B$ and $G_\pm$ (where $\Omega_r$ and $\Omega_m$ become
smaller and smaller). In particular, for the point $B$ we obtain
$2+\alpha_3+\alpha_4<0$, for the point $G_+$ we acquire $\alpha_3>0$,
$0<\alpha_4<\frac{\alpha_3^2}{8}$ and for the point $G_-$ we obtain
$\alpha_3<0$  and $0<\alpha_4<\frac{\alpha_3^2}{8}$.

In order to present the cosmological behavior more transparently, we
numerically evolve the autonomous system (\ref{eqr})-(\ref{eqxi})
for the  the parameter choice $\alpha_3=1$,$\alpha_4=1$ and $\omega=0.01$,
and in Fig. \ref{Fig1} we depict the corresponding phase-space behavior in
the $\Omega_{DE}$-$\Omega_m$ plane.

\section{Cosmological evolution}
\label{Cosmological evolution}

In the previous section we performed a dynamical analysis of quasi-Dilaton
non-linear massive gravity, that is we focused on its asymptotic behavior
at late times, that is on solutions that are going to attract the universe
independently of the initial conditions and of the specific cosmological
evolution towards them. In this section we investigate the behavior of the universe at 
intermediate times, which obviously does depend on the
initial conditions, but it can be very interesting. In particular, we are
interested in obtaining a evolution of the universe in agreement with the
observed epoch sequence. Furthermore, we examine the possibility of
obtaining bouncing behavior. In the following subsections we discuss these
two cosmological behaviors separately.

\subsection{Epoch sequence and late-time acceleration }

Let us first examine the universe evolution, focusing on the various
density parameters and the dark-energy equation of state. Observing the
Friedmann equation (\ref{eq:Fried}), as well as the evolution equations for
the density parameters and for $\xi$ (\ref{eqr})-(\ref{eqxi}), we deduce
that the observed post-inflationary thermal history of the universe can be
easily obtained by suitably choosing the model parameters.

\begin{figure}[!]
\begin{center}
\mbox{\epsfig{figure=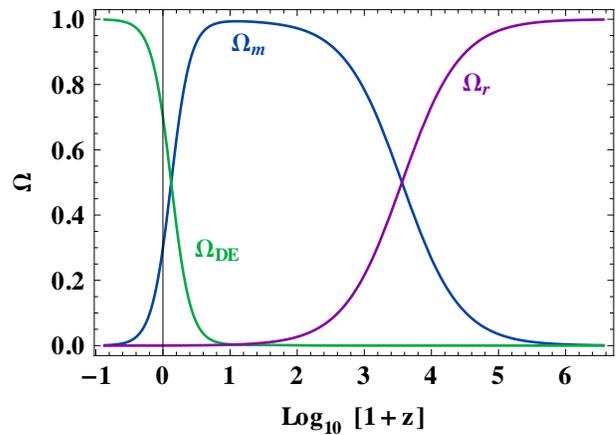,width=8.6cm,angle=0}}
\caption{{\it Cosmological evolution of the radiation ($\Omega_r$), matter
($\Omega_m$) and dark energy
($\Omega_{DE}\equiv\Omega_\Lambda+\Omega_\sigma$) density parameters, as a
function of the redshift $z$, for the parameter choice $\alpha_3=1$,
$\alpha_4=0.115$ and $\omega=0.01$. In this specific example the observed
universe epoch sequence is reproduced.}}
 \label{density}
\end{center}
\end{figure}

In order to present this behavior more transparently, we numerically evolve
the system for $\alpha_3=1$, $\alpha_4=0.115$ and $\omega=0.01$, starting
with initial conditions corresponding to $\Omega_r\approx1$, and imposing
the current values to be $\Omega_m\approx0.28$ and
$\Omega_{DE}\approx0.72$. In Fig. \ref{density} we depict the resulting
evolution for the density parameters using, instead of the scale factor,
the redshift $z$ as the independent variable ($1 + z =a_0/a$ with $a_0=1$
the present scale-factor value). As we observe, the thermal history of the
universe can be reproduced, namely we obtain the successive sequence of
radiation, matter and dark energy epochs.

\begin{figure}[!]
\begin{center}
\mbox{\epsfig{figure=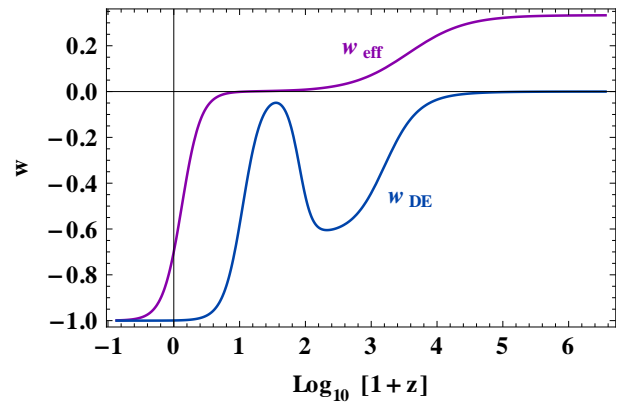,width=8.1cm,angle=0}}
\caption{{\it
Cosmological evolution of the total ($w_{eff}$) and of the dark-energy
($w_{DE}$) equation-of-state parameters, as a
function of the redshift $z$, for the parameter choice $\alpha_3=1$,
$\alpha_4=0.115$ and $\omega=0.01$. In this specific example, dark energy
starts from a dust-like behavior, resulting in a cosmological-constant-like
one at the current epoch, which drives the observed universe
acceleration.}} \label{eos}
\end{center}
\end{figure}

Similarly, in Fig. \ref{eos} we present the corresponding evolution of the
total equation-of-state parameter $w_{eff}$ as well as of the dark-energy
one $w_{DE}$. As we observe, dark energy starts from a dust-like behavior,
resulting in a cosmological-constant-like one at the current universe,
driving the observed acceleration. Note that, as expected, in the future,
where dark energy completely dominates, both $w_{eff}$ and $w_{DE}$
coincide at the cosmological constant value.

\subsection{Bouncing and turnaround solutions}

In this subsection we examine the bounce realization in the scenario at
hand. Bouncing cosmologies in general have gained significant interest,
since they offer an alternative paradigm free of the ``Big-Bang
singularity'', as well as of the horizon, flatness and monopole problems
\cite{Novello:2008ra}. The basic condition for the bounce is the Hubble
parameter to change sign from negative to positive, while its
time-derivative to be positive. Additionally, when the Hubble parameter
changes from positive to negative,  while its time-derivative is
negative, we have the realization of a cosmological turnaround. Such a
behavior implies the Null Energy Condition (NEC) violation around the
bounce (or the turnaround), which is nontrivial in the context of General
Relativity \cite{Nojiri:2013ru}, leading in general to ghost degrees of
freedom \cite{Carroll:2003st,Cline:2003gs}. However, a
safe bounce evolution, free of ghosts and instabilities, is possible in
various modified gravity constructions
\cite{Cai:2011bs,Capozziello:2011et}. Clearly, observing the Friedmann
equation (\ref{eq:Fried}) of the present scenario of quasi-Dilaton
non-linear massive gravity, we deduce that the basic bounce (or turnaround)
realization
condition can be easily fulfilled.

In order to quantify the above discussion, and assuming standard dust
matter and
standard radiation, we rewrite the Friedmann equation
(\ref{eq:Fried}) in the following form:
\begin{align}
\label{FR1aux}
\frac{H^2}{H_0^2}=\frac{\Omega_{m0}a^{-3}+\Omega_{r,0}a^{-4}-\frac{m^2}{
H_0^2}G_1}{1- \frac{\omega}{6}
\Bigl(1-4\frac{G_2}{\xi G_2'}\Bigr)^2},
\end{align}
where $\Omega_{m0}$ and $\Omega_{r,0}$ are the values of the
corresponding density parameters at present ($a_0=1$), and $H_0$ the
current Hubble parameter. Furthermore, since the constraint equation
(\ref{eq:Constraint}) provides $\xi$ as a function of the scale factor, we
can finally write (\ref{FR1aux}) as
\begin{align}
\label{Va}
\frac{H^2}{H_0^2}=V(a),
\end{align}
with $C,\alpha_3,\alpha_4,\omega,m$ and $\Omega_{m0},\Omega_{r,0}$ as
parameters. In summary, examining the behavior of the above known $V(a)$,
we can find possible bouncing points at $a=a_B$ (when $V(a_B)=0$ and
$H_0^{-1}\frac{d}{dt}\sqrt{V(a_B)}>0$), or possible turnaround points
at $a=a_T$
(when $V(a_T)=0$ and
$ H_0^{-1}\frac{d}{dt}\sqrt{V(a_T)}<0$). This procedure is
straightforward, and thus
we deduce that both bounce and turnaround are possible in the scenario of
quasi-Dilaton non-linear massive gravity. Although one can apply it in
full generality, obtaining exact results, since the involved expressions
for $a_B$ and $a_T$ are lengthy in the following we explicitly apply it in
a specific simple example, where simple expressions can be easily obtained.

First of all, since a bounce occurs in the early universe, that is at
small scale factors ($a\ll a_0=1$), from (\ref{eq:Constraint}) we deduce
that at the bounce $\xi\gg1$. On the other hand, since a turnaround occurs
in the late universe, that is at $a\gg a_0=1$, from (\ref{eq:Constraint})
we deduce that at the turnaround $G_2(\xi)=0$.

Let us for simplicity investigate the simplest case $\alpha_3=\alpha_4=0$.
At the bounce region from (\ref{G1def}) we obtain $G_1\approx \xi^2$ and
 $G_2 \approx -\xi^2 =C/a^4$, which implies that $C<0$. Therefore, from
(\ref{FR1aux}) we deduce that
\begin{equation}
\frac{H^2}{H_0^2}\approx\frac{\Omega_{m0}\left(1-\frac{a_B}{
a}\right)a^{-3}}{1-\omega/6},
\end{equation}
with
\begin{align}
a_B=-\frac{\Omega_{r,0}+\frac{C m^2}{H_0^2}}{\Omega_{m0}},
\label{ab1}
 \end{align}
and therefore
 \begin{equation}
 \frac{\dot{H}}{H_0^2}\approx-\frac{3\Omega_{m0}a^{-3}\left(3-4\frac{a_B}{
a}\right)}{6-\omega}.
 \label{hd1}
\end{equation}
From the above relations we easily acquire that $H(a_B)=0$ and moreover for
$\omega<6$, which is always the case in interesting cosmology as we
discussed in section \ref{Dynamicalbehavior}, we have  $\dot{H}(a_B)>0$.
Thus, we immediately see that the bounce is obtained at $a=a_B$. Finally,
note that obviously
we require $a_B>0$, which gives an additional constraint  on the
parameters,
namely $Cm^2/H_0^2<-\Omega_{r,0}$.

Let us now see whether a turnaround can occur in this simplest case
$\alpha_3=\alpha_4=0$. For $a\gg1$, that is for $G_2=0$, we have $\xi=0$
or $\xi=1$. Obviously, since $\xi$ varies, the above conditions are quite
difficult to be obtained, and thus contrary to the bounce, the
turnaround is harder to be acquired. In the $\xi=0$ case from (\ref{G1def})
we acquire $G_1=2$,
and therefore from (\ref{FR1aux})  we obtain
\begin{equation}
\frac{H^2}{H_0^2}\approx\frac{\Omega_{m0}\left(a^{-3}-a_T^{-3}\right)}{1-
\frac{3\omega}{2}}
\end{equation}
with
\begin{equation}
 a_T=\(\frac{H_0^2\Omega_{m0}}{2m^2}\)^{1/3},
\end{equation}
where for simplicity we neglected the radiation term since at large scale
factors the matter term will dominate in general. Thus,
\begin{equation}
 \frac{\dot{H}}{H_0^2}\approx-\frac{\frac{3}{2}\Omega_{m0}a^{-3}}{
1-\frac{
3\omega}{2} }.
\end{equation}
From these relations we deduce that
$H(a_T)=0$, while $\dot{H}(a_T)<0$ for $\omega<2/3$, which is the
turnaround condition (note that in this case $a_T>0$ always).
Concerning the second case $\xi=1$, from (\ref{G1def}) we acquire
$G_1=0$, and thus
$ H^2 \propto \Omega_{m0}a^{-3}+\Omega_{r,0}a^{-4} $ which has no roots,
and hence we deduce that in this case the turnaround is impossible.

\begin{figure}
\includegraphics[scale=.7]{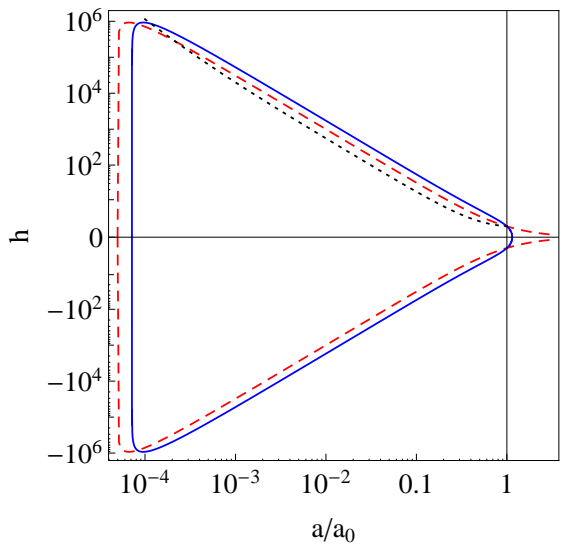}
\caption{{\it{The rescaled Hubble function $h=H/H_0$ as a function of
$a/a_0$, for
$\alpha_3=\alpha_4=0$, $\Omega_m=0.3$, $\Omega_r=10^{-4}$, $C=-1.1 \times
10^{-4}$ and $m=H_0$. In this case we obtain two evolution branches, one
corresponding to a
cyclic Universe (blue line) and a second branch corresponding to a Universe
with a past bounce (red dashed curve). The dotted line corresponds to $\Lambda$CDM
scenario with the same cosmological parameters.}}}
\label{fig:bounce}
\end{figure}

Finally, note that for $\omega<2/3$ and $Cm^2/H_0^2<-\Omega_{r,0}$, in the
above simplest example we obtain both a bounce and a turnaround, which is the
realization of cyclic cosmology \cite{Novello:2008ra}. In 
Fig. \ref{fig:bounce}, we present such a behavior, by numerically evolving
the exact cosmological equations for parameter choices that satisfy the above
conditions.

One can straightforwardly perform the above analysis for general
$\alpha_3$ and $\alpha_4$, extracting the corresponding conditions for a
bounce and turnaround, or for their simultaneous realization, that is for
cyclic cosmology. In summary, as we can see, in the scenario of
quasi-Dilaton non-linear massive gravity the above alternative
cosmological evolutions can be easily obtained, in agreement with the
general case of extended non-linear massive gravity \cite{Cai:2012ag}.

\section{Observational Constraints}
\label{observations}

Having analyzed the cosmological behavior in the scenario of quasi-Dilaton
non-linear massive gravity, we now proceed to investigate the observational
constraints on the model parameters. We use observational data from Type
Ia Supernovae (SNIa), Baryon Acoustic Oscillations (BAO), and Cosmic
Microwave Background (CMB), and as usual we work in the standard units
suitable for observational comparisons, namely setting  $8\pi G=c=1$.
\begin{figure}[ht]
\begin{center}
\mbox{\epsfig{figure=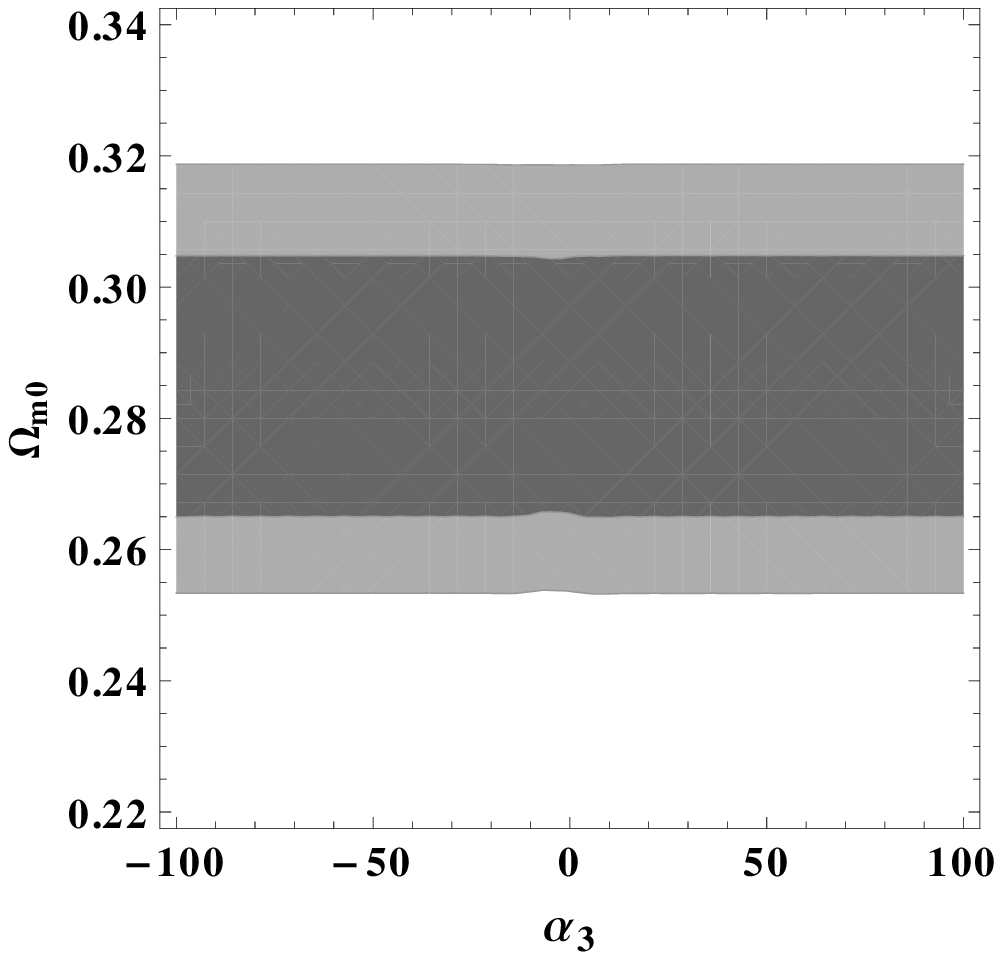,width=6.8cm,angle=0}}
\mbox{\epsfig{figure=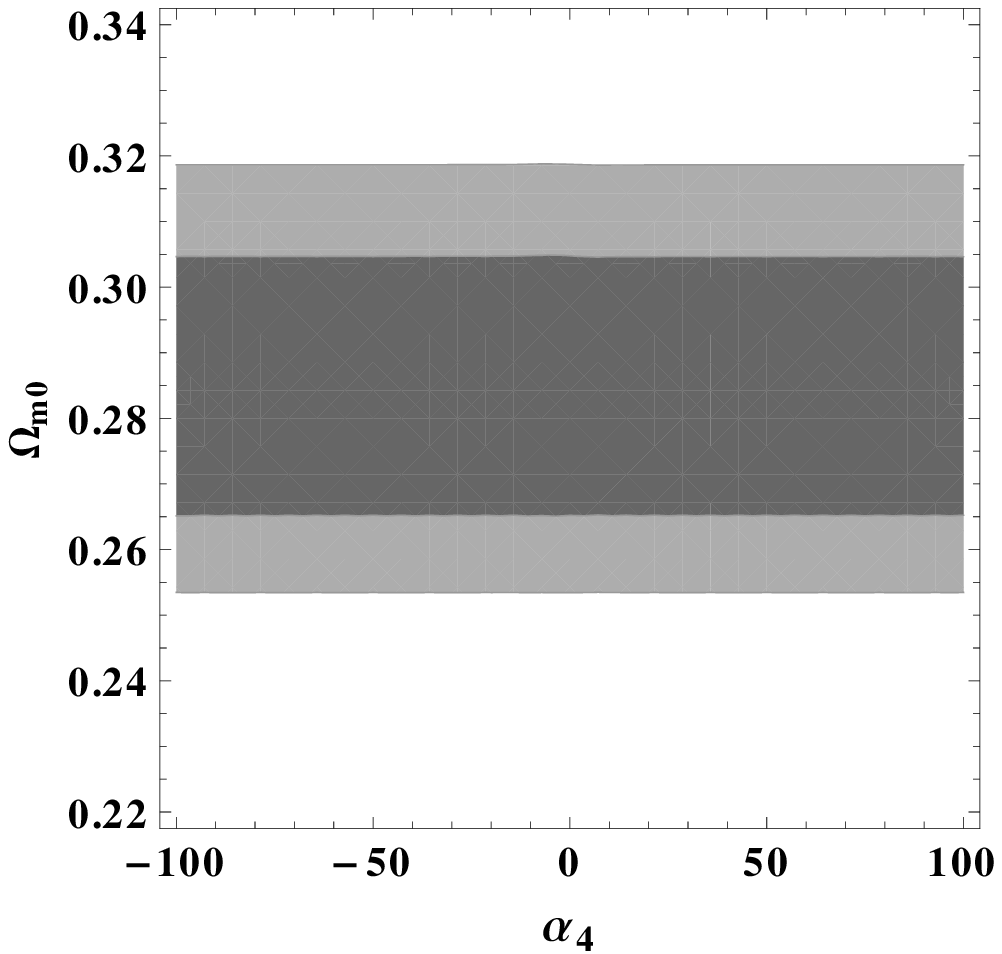,width=6.8cm,angle=0}}
\caption{{\it Top: The 1$\sigma$ (dark) and 2$\sigma$ (light)
likelihood
contours in the $\alpha_3-\Omega_{m0}$ plane, using SNIa, BAO and CMB
data, for $\alpha_4=1$. Bottom: The 1$\sigma$
(dark) and 2$\sigma$ (light) likelihood contours in the
$\alpha_4-\Omega_{m0}$ plane using SNIa, BAO and CMB
data, for $\alpha_3=1$. Both graphs are for
$\omega=0.01$.}} \label{alpha}
\end{center}
\end{figure}

In order to perform the analysis, and assuming standard dust matter and
standard radiation ($w_r=1/3$), we re-write the Friedmann equation
(\ref{eq:Fried}) in the form of (\ref{FR1aux}), and we use the
redshift $z$ instead of the scale factor. Thus, the scenario at hand has
the parameters $C,\alpha_3,\alpha_4,\omega,m$ and $\Omega_{m0}$ or
$\Omega_{DE0}$ (we fix $H_0$ by its 7-year WMAP best-fit value
\cite{Komatsu:2010fb}). The details of the  combined
SNIa+CMB+BAO  analysis are given in the
Appendix, and here we present
the constructed likelihood contours, discussing their structure.

Let us first examine the constraints on $\alpha_3$ and $\alpha_4$. In Fig.
\ref{alpha} we depict the $1\sigma$ and
$2\sigma$   $\alpha_3-\Omega_{m0}$ and  $\alpha_4-\Omega_{m0}$
contours respectively.
As we observe, the parameters $\alpha_3$ and $\alpha_4$ are not
constrained by the observational data, that is they only slightly affect
the cosmological evolution. This behavior verifies the results obtained
in section \ref{Dynamicalbehavior}, where the stable late-time solutions
were found to be independent of $\alpha_3$,$\alpha_4$, as well as those
obtained in section \ref{Cosmological evolution}, where at intermediate
times different values of $\alpha_3$,$\alpha_4$ lead to nearly the same
cosmology. This is in agreement with the analysis of \cite{Gong:2012ny}
for the simple non-linear massive gravity. Finally, concerning
$\Omega_{m0}$ we obtain a best fit value approximately at $0.284$
for both $\alpha_3$ and $\alpha_4$.

\begin{figure}[ht]
\begin{center}
\mbox{\epsfig{figure=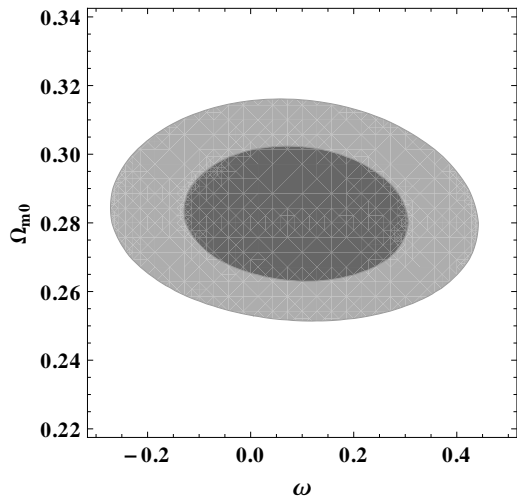,width=6.8cm,angle=0}}
\caption{{\it The 1$\sigma$ (dark) and 2$\sigma$ (light) likelihood
contours
in the $\omega-\Omega_{m0}$ plane using SNIa, BAO and CMB
data, for $\alpha_3=1$ and $\alpha_4=0.115$.
}}
\label{omega}
\end{center}
\end{figure}

In order to extract the constraints on the quasi-dilaton parameter
$\omega$, and knowing the above results on the insignificant role of
$\alpha_3$ and $\alpha_4$, we fix them to $\alpha_3=1$,$\alpha_4= 0.115$
(alternatively we could consider them as free parameters, marginalizing
with flat priors, but the results would be almost the same). In Fig.
\ref{omega} we depict the $1\sigma$ and $2\sigma$ likelihood contours in
the $\omega-\Omega_{m0}$ plane.
As we observe $\omega$ is significantly constrained by observations, with
its best-fit value being $\omega=0.0878$. Note also that negative $\omega$
values are also allowed, leading the dilaton to behave as a phantom field.
This strong constraining behavior verifies the results of the dynamical
analysis of section \ref{Dynamicalbehavior}, where in the case of
interesting late-time cosmology $\omega$ was bounded. Finally, we mention
that apart from the above late-time observational data, we could use Big
Bang Nucleosynthesis arguments  \cite{{BBNrefs1}} in order to impose
constraints on
 $\Omega_{DE}$, that is to $\omega$, however these would be weaker than
the ones obtained above.

\begin{figure}[ht]
\begin{center}
\mbox{\epsfig{figure=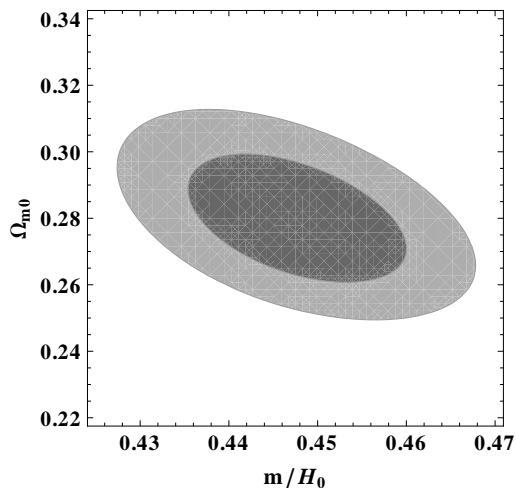,width=6.8cm,angle=0}}
\caption{{\it The 1$\sigma$ (dark) and 2$\sigma$ (light) likelihood
contours
in the $m/H_0-\Omega_{m0}$ plane using SNIa, BAO and CMB
data, for $\alpha_3=1$ and $\alpha_4=0.115$.
}}
\label{m-H0}
\end{center}
\end{figure}

Finally, we examine the constraints on the graviton mass $m$, and in 
Fig. \ref{m-H0} we depict the corresponding likelihood contours in
relation to $\Omega_{m0}$. From this figure we observe that the ratio
$m/H_0$ is highly constrained, and in particular we verify the expected
result  $m\sim H_0$, although with $m$ being a bit smaller than $H_0$.
In particular, the best-fit values are $m/H_0=0.44775$ and
$\Omega_{m0}=0.2794$.

\section{Conclusions}
\label{Conclusions}

In this work we investigated the cosmological dynamics of the
quasi-Dilaton non-linear massive gravity. In such a gravitational
construction, one uses the usual non-linear massive gravity
\cite{deRham:2010ik,deRham:2010kj}, which is free of ghosts and couples it with a dilaton field
\cite{D'Amico:2012zv}.  Apart from the matter and radiation sectors
in this scenario we obtain an effective dark energy, constituted
from the massive graviton and the dilaton contributions. The
corresponding cosmological behaviors proves to be very interesting.

First we performed a detailed dynamical analysis, and we extracted the
stable late-time solutions, that is solutions that always attract the
universe at late times. Furthermore, in each of these states we have
calculated the corresponding observables, such as the dark-energy and
matter density parameters, the dark-energy equation-of-state parameter, and
the deceleration parameter. We
showed that the scenario at hand gives rise to a dark-energy
dominated accelerating universe where dark energy behaves like
cosmological constant.

Leaving aside the good asymptotic behavior at late times, the
cosmological behavior of quasi-Dilaton non-linear massive gravity has
also interesting  intermediate epochs {\it a la} thermal history. We
have shown that the observed history of the universe can be
easily obtained in a parameter region, that is the expected sequence
of radiation, matter, and dark energy epochs. As a bonus, we found
conditions on the parameters that might lead to bouncing, turnaround
or cyclic behavior which is an interesting feature of the scenario
under consideration.

Last but not least, we have imposed observational constraints on the
model parameters. We carried out model-fitting using data from Type Ia
Supernovae (SNIa), Baryon Acoustic Oscillations (BAO), and Cosmic
Microwave Background (CMB) observations. Constructing the
corresponding contour plots we deduced that the usual $\alpha_3$ and
$\alpha_4$ parameters of non-linear massive gravity are not
constrained by current data at the background level. However, the
dilaton-coupling parameter is significantly constrained at a narrow
window around zero. Finally, as expected, the graviton mass is found to be of
the same order of the present value of the Hubble parameter.

From the aforesaid, we find that  cosmology of the quasi-Dilaton
non-linear massive gravity is  interesting and is in agreement with
the observed universe behavior. In our opinion, it can be a good
candidate for the description of universe. However, in order to
obtain additional indications of the validity and  consistency with
observations, of the scenario at hand, apart from the background
analysis that was the base of this work, we should proceed to a
detailed examination of the perturbations. In particular, one would
need to investigate the perturbation-related observables such as the
spectral index, the tensor-to-scalar ratio, and the non-gaussianity
estimators. Such an analysis lies beyond the goal of the present
work and it is left for future investigations. Last but not least we
should comment on the issue of superluminality in the model under
consideration. As pointed out by Deser and Waldron \cite{Deser:2012qx}, superluminal
behavior is an essential feature of dRGT. In the quasi-Dilaton
model, in the decoupling limit, we have a standard scalar field
$\sigma$ in addition to the galileon which is plagued with
superluminality. Since $\sigma$ has canonical kinetic term, it looks
less likely that it would affect the superluminal behavior of the
model. However, the problem requires separate investigations to
address the issue which we defer to our future work.

\begin{acknowledgments}
The authors wish to thank G. Leon for useful comments. M.W.H
acknowledges the funding from CSIR, Govt of India. MS thanks
G.~Gabadadze for a brief and fruitful discussion on the subject
during Kyoto meeting. MS is also thankful to K. Bamba and S. Nojiri
for useful discussions on the related theme. The research of
E.N.S. is implemented within the framework of the Operational
Program ``Education and Lifelong Learning'' (Actions Beneficiary:
General Secretariat for Research and Technology), and is co-financed
by the European Social Fund (ESF) and the Greek State. 
\end{acknowledgments}

\appendix*

\section{Observational data and constraints}
\label{Observ}

Here we briefly review the sources of observational constraints used in
this manuscript, namely Type Ia Supernovae constraints, Baryon Acoustic
Oscillations (BAO) and Cosmic Microwave Background (CMB). The total
$\chi^2$ defined as
\begin{align}
\chi^2=\chi_{SN}^2+\chi_{BAO}^2+\chi_{CMB}^2,
\end{align}
where the individual $\chi^2_i$ for every data set is calculated as
follows.   \\

{\it{a. Type Ia Supernovae constraints}}\\

\begin{table*}[!]
\begin{center}
\begin{tabular}{|c||c|c|c|c|c|c|}
\hline
 $z_{BAO}$  & 0.106  & 0.2 & 0.35 & 0.44 & 0.6 & 0.73\\
\hline \hline
 $\frac{d_A(z_\star)}{D_V(Z_{BAO})}$ &  $30.95 \pm 1.46$ & $17.55 \pm 0.60$
& $10.11 \pm 0.37$ & $8.44 \pm 0.67$ & $6.69 \pm 0.33$ & $5.45 \pm 0.31$
\\
\hline
\end{tabular}
\caption{Values of $\frac{d_A(z_\star)}{D_V(Z_{BAO})}$ for different values
of $z_{BAO}$.}
\label{baodata}
\end{center}
\end{table*}
In order to incorporate Type Ia constraints  we use the Union2.1 data
compilation \cite{Suzuki:2011hu} of 580 data points. The relevant
observable is the distance modulus $\mu$ which is defined as  $\mu=m - M=5
\log D_L+\mu_0$,
where $m$ and $M$ are the apparent and absolute
magnitudes of the Supernovae, $D_L(z)$ is the luminosity distance
$D_L(z)=(1+z)
\int_0^z\frac{H_0dz'}{H(z')}$ and
$\mu_0=5 \log\left(\frac{H_0^{-1}}{M_{pc}}\right)+2
5$ is a nuisance parameter that should be marginalized. The
corresponding $\chi^2$ writes as
\begin{align}
\chi_{SN}^2(\mu_0,\theta)=\sum_{i=1}^{580}
\frac{\left[\mu_{th}(z_i,\mu_0,\theta)-\mu_{obs}(z_i)\right]^2}{
\sigma_\mu(z_i)^2},
\end{align}
where $\mu_{obs}$ denotes the observed distance modulus while $\mu_{th}$
the theoretical one, and $\sigma_{\mu}$ is the uncertainty in the distance
modulus. Additionally, $\theta$ denotes any parameter of the specific
model at hand. Finally, marginalizing $\mu_0$ following
\cite{Lazkoz:2005sp} we obtain
\begin{align}
\chi_{SN}^2(\theta)=A(\theta)-\frac{B(\theta)^2}{C(\theta)},
\end{align}
with
\begin{align}
&A(\theta) =\sum_{i=1}^{580}
\frac{\left[\mu_{th}(z_i,\mu_0=0,\theta)-\mu_{obs}(z_i)\right]^2}{
\sigma_\mu(z_i)^2}
\end{align}
\begin{align}
&B(\theta) =\sum_{i=1}^{580}
\frac{\mu_{th}(z_i,\mu_0=0,\theta)-\mu_{obs}(z_i)}{\sigma_\mu(z_i)^2}\\
&C(\theta) =\sum_{i=1}^{580} \frac{1}{\sigma_\mu(z_i)^2}.
\end{align}

{\it{b. Baryon Acoustic Oscillation constraints}}\\

We use BAO data from
\cite{Blake:2011en,Percival:2009xn,Beutler:2011hx,Jarosik:2010iu},
that is of $\frac{d_A(z_\star)}{D_V(Z_{BAO})}$,
where $z_\star$ is the decoupling time given by $z_\star \approx 1091$, the
co-moving angular-diameter distance $d_A(z)=\int_0^z \frac{dz'}{H(z')}$
and
$D_V(z)=\left(d_A(z)^2\frac{z}{H(z)}\right)^{\frac{1}{3}}$
is the dilation scale \cite{Eisenstein:2005su}. th required data
are depicted in Table $\ref{baodata}$.

In order to calculate $\chi_{BAO}^2$ for BAO data we follow the procedure
described in \cite{Giostri:2012ek}, where it is defined as,
\begin{equation}
 \chi_{BAO}^2=X_{BAO}^T C_{BAO}^{-1} X_{BAO},
\end{equation}
with
\begin{equation}
X_{BAO}=\left( \begin{array}{c}
        \frac{d_A(z_\star)}{D_V(0.106)} - 30.95 \\
        \frac{d_A(z_\star)}{D_V(0.2)} - 17.55 \\
        \frac{d_A(z_\star)}{D_V(0.35)} - 10.11 \\
        \frac{d_A(z_\star)}{D_V(0.44)} - 8.44 \\
        \frac{d_A(z_\star)}{D_V(0.6)} - 6.69 \\
        \frac{d_A(z_\star)}{D_V(0.73)} - 5.45
        \end{array} \right)
\end{equation}
and the inverse covariance matrix reads as
\begin{widetext}
\begin{align}
C^{-1}=\left(
\begin{array}{cccccc}
 0.48435 & -0.101383 & -0.164945 & -0.0305703 & -0.097874 & -0.106738 \\
 -0.101383 & 3.2882 & -2.45497 & -0.0787898 & -0.252254 & -0.2751 \\
 -0.164945 & -2.45499 & 9.55916 & -0.128187 & -0.410404 & -0.447574 \\
 -0.0305703 & -0.0787898 & -0.128187 & 2.78728 & -2.75632 & 1.16437 \\
 -0.097874 & -0.252254 & -0.410404 & -2.75632 & 14.9245 & -7.32441 \\
 -0.106738 & -0.2751 & -0.447574 & 1.16437 & -7.32441 & 14.5022
\end{array}
\right).
\end{align}
\end{widetext}

{\it{c. CMB constraints}}\\

We use the CMB shift parameter
\begin{align}
R=H_0 \sqrt{\Omega_{m0}}
\int_0^{1089}\frac{dz'}{H(z')}
\end{align}
 following \cite{Komatsu:2010fb}.
The corresponding $\chi_{CMB}^2$ is defined as,
\begin{align}
 \chi_{CMB}^2(\theta)=\frac{(R(\theta)-R_0)^2}{\sigma^2},
\end{align}
with $R_0=1.725 \pm 0.018$ and $R(\theta)$ \cite{Komatsu:2010fb}. \\

\end{document}